\documentclass[conference]{IEEEtran}
\IEEEoverridecommandlockouts

\usepackage{cite}
\usepackage{amsmath,amssymb,amsfonts}
\usepackage{amsthm}
\newtheorem{theorem}{Theorem}
\usepackage{algorithm}
\usepackage{algpseudocode}
\usepackage{graphicx}
\usepackage{epstopdf}
\usepackage{epsfig}
\usepackage{textcomp}
\usepackage{xcolor}
\usepackage[font=normalsize]{caption}

\begingroup\makeatletter\def\f@size{8}\check@mathfonts
\def\maketag@@@#1{\hbox{\m@th\normalsize\normalfont#1}}
\endgroup

\begingroup
\allowdisplaybreaks

\usepackage{geometry}
\def\BibTeX{{\rm B\kern-.05em{\sc i\kern-.025em b}\kern-.08em
    T\kern-.1667em\lower.7ex\hbox{E}\kern-.125emX}}
    
\begin{document}
\title{Antenna Position Optimization for Movable Antenna-Empowered Near-Field Sensing}

\author{\IEEEauthorblockN{Yushen Wang\IEEEauthorrefmark{1}, Weidong Mei\IEEEauthorrefmark{2}, Xin Wei\IEEEauthorrefmark{2}, Boyu Ning\IEEEauthorrefmark{2}, and Zhi Chen\IEEEauthorrefmark{2}}
		\IEEEauthorblockA{\IEEEauthorrefmark{1}Glasgow College,
		}
		\IEEEauthorblockA{\IEEEauthorrefmark{2}National Key Laboratory of Wireless Communications,}
		University of Electronic Science and Technology of China, Chengdu, China.\\
		
		Email: 2022190903007@std.uestc.edu.cn, wmei@uestc.edu.cn, xinwei@std.uestc.edu.cn,\\ boydning@outlook.com, chenzhi@uestc.edu.cn
        \thanks{This work was supported in part by the National Key Research and Development Program of China under Grants 2024YFB2907900 and 2024YFE0200400.}}

\maketitle

\begin{abstract}
Movable antennas (MAs) show great promise for enhancing the sensing capabilities of future sixth-generation (6G) networks. With the growing prevalence of near-field propagation at ultra-high frequencies, this paper focuses on the application of MAs for near-field sensing to jointly estimate the angle and distance information of a target. First, to gain essential insights into MA-enhanced near-field sensing, we investigate two simplified cases with only the spatial angle-of-arrival (AoA) or distance estimation, respectively, assuming that the other information is already known. We derive the worst-case Cramer-Rao bounds (CRBs) on the mean square errors (MSEs) of the AoA estimation and the distance estimation via the multiple signal classification (MUSIC) algorithm in these two cases. Then, we jointly optimize the positions of the MAs within a linear array to minimize these CRBs and derive their closed-form solutions, which yield an identical array geometry to MA-aided far-field sensing. Furthermore, we proceed to the more challenging case with the joint AoA and distance estimation and derive the worst-case CRB under the two-dimensional (2D) MUSIC algorithm. The corresponding CRB minimization problem is efficiently solved by adopting a discrete sampling-based approach. Numerical results demonstrate that the proposed MA-enhanced near-field sensing significantly outperforms conventional sensing with fixed-position antennas (FPAs). Moreover, the joint angle and distance estimation results in a different array geometry from that in the individual estimation of angle or distance.
\end{abstract}

\section{Introduction}
In future sixth-generation (6G) wireless systems, significant advancement in their communication and sensing capabilities is envisioned\cite{1, 2}. A multitude of research efforts have been dedicated to achieving ultra-high transmission rates and precise information acquisition from the physical environment, driven by the demands of emerging applications such as smart healthcare, vehicle-to-everything (V2X) and virtual reality (VR) \cite{3, 4}. However, existing communication and/or sensing systems typically deploy fixed-position antennas (FPAs) at the transmitter/receiver (Tx/Rx), which cannot fully exploit the spatial degrees of freedom (DoFs).

To tackle this inflexibility, movable antenna (MA) technology has received increasing attention in wireless networks. On the one hand, the MAs can move continuously within a confined region at the Tx/Rx to adaptively configure the channel conditions for different purposes in wireless communications, e.g., spatial multiplexing and interference mitigation \cite{5, 6, 7}. As such, MA position optimization for communications has been widely investigated under various system setups, including multiple-input single-/multiple-output (MISO/MIMO) system \cite{8, 9}, physical-layer security \cite{10}, over-the-air computation \cite{11}, cognitive radio \cite{12}, etc. On the other hand, MA technology also shows great potential to empower wireless sensing, which is the focus of this work. By leveraging a broader region at the Tx/Rx for antenna movement, the aperture of MA arrays can be enlarged compared to FPA arrays, which increases their angle and distance estimation resolution \cite{6}. In \cite{13, 14, 15, 16}, the authors have investigated the MA position optimization problems for sensing or integrated sensing and communications (ISAC) and characterized the Cramer-Rao bound (CRB) in the estimation as well as its trade-off with the communication performance. However, all of the above works only focused on far-field sensing. To enhance the design flexibility for MAs, MA-aided wireless networks typically feature large antenna movement regions. Moreover, current wireless communication systems are expected to migrate to higher frequency bands (e.g., Terahertz (THz) bands) in future to achieve broader bandwidth. The combination of larger antenna apertures and higher frequency bands necessitates the use of the near-field spherical-wave model for communication or sensing \cite{17, 18, 19}. However, to the best of our knowledge, there is no existing work focusing on MA-enhanced near-field sensing so far.
\begin{figure}[t]
    \centering
    \includegraphics[scale=0.35]{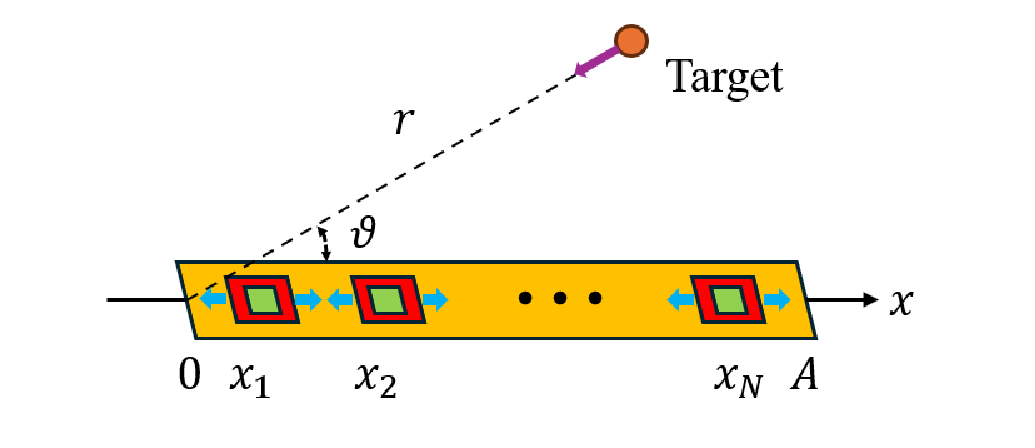}
    \caption{\small MA-enhanced near-field target sensing.}
    \label{Fig1}
\end{figure}
To fill in this gap, we focus on MA-enhanced near-field sensing in this paper, aiming to estimate the angle or/and distance information of a target with the aid of a linear MA array, as shown in Fig. \ref{Fig1}. To gain essential insights, we first investigate the individual estimation of the spatial angle-of-arrival (AoA) or distance via the multiple signal classification (MUSIC) algorithm under the assumption that the other parameter is already known. The worst-case CRB on the mean square error (MSE) of the AoA/distance estimation is derived and minimized by optimizing the antenna position vector (APV). The closed-form solutions to the CRB minimization problems in these two cases are derived, which yield an identical array geometry to that in far-field sensing. Furthermore, we proceed to the general case with the joint AoA and distance estimation and derive the worst-case sum of the CRBs under the two-dimensional (2D) MUSIC algorithm. To resolve the more challenging CRB minimization problem, a discrete sampling-based approach is proposed by discretizing the movement region into a multitude of sampling points and sequentially updating the positions of the MAs until convergence. Numerical results demonstrate the efficacy of the proposed sensing scheme compared to conventional FPAs and also show a different array geometry for the joint AoA and distance estimation versus the individual estimation.

\textit{Notations}: Boldface lower and upper cases are used to represent vectors and matrices, respectively. The conjugate of a complex is denoted by $(\cdot)^*$. The transpose and conjugate transpose of a matrix/vector are denoted by $(\cdot)^\top$ and $(\cdot)^\mathrm{H}$, respectively. The sets of $N_1 \times N_2$ dimensional real and complex matrices are denoted by $\mathbb{R}^{N_1 \times N_2}$ and $\mathbb{C}^{N_1 \times N_2}$, respectively. The $l_2$-norm of vector $\boldsymbol{v}$ is denoted by $\lVert\boldsymbol{v}\rVert$. For a real number $c$, $\lvert c \rvert$ denotes its absolute value and $\lfloor c \rfloor$ denotes the maximum integer that is no larger than $c$. The expectation is denoted by $\mathbb{E}\{\cdot\}$. $\boldsymbol{I}_N$ denotes the $N$-dimensional identity matrix.

\section{System Model}

As illustrated in Fig. \ref{Fig1}, we consider a 1D near-field wireless sensing system with \( N \) MAs to estimate the angular and spatial parameters of a target. The positions of the MAs can be flexibly adjusted within a given 1D line segment of length \( A \). Denote the position of the \( n \)-th MA (\( n \in {\cal N} \triangleq \{1, 2, \ldots, N\} \)) by \( x_n \in [0, A] \), and the APV of all \( N \) MAs by $\boldsymbol{x} \triangleq [x_1, x_2, \ldots, x_N]^\top \in \mathbb{R}^N$. Without loss of generality, we assume that \( 0 \leq x_1 < x_2 < \cdots < x_N \leq A \). Therefore, the effective aperture of the MA array can be represented as $D = x_N - x_1$. We assume that the target is located in the near-field region of the linear array but outside its reactive region, which means that the distance between the target and any position within the MA array is between the Fresnel distance and the Rayleigh distance, which are respectively given by $R_{FS} \triangleq \frac{A}{2}(\frac{A}{\lambda})^{\frac{1}{3}}$ \cite{20} and $R_{RL} \triangleq \frac{2A^2}{\lambda}$ \cite{19}, where $\lambda$ is the signal wavelength.
 
In the sensing process, we consider that the MA array emits a sensing signal and receives the echoes reflected from the target across \textit{T} snapshots, during which the target is assumed to remain static \cite{18, 21}. To characterize the near-field channel model from the antenna array to the target, we adopt a uniform spherical wave (USW) channel model in \cite{19}, where the path loss of the channel coefficient with respect to (w.r.t.) all MAs is identical while the phase varies over them. As depicted in Fig. \ref{Fig1}, we denote the physical steering angle from the origin to the target as $\theta$, while the spatial AoA is defined as $u \triangleq \cos\theta \in [0, 1)$\footnote{We exclude the case with $u < 0$, as it is equivalent to placing the array in the other side, which yields the same sensing result as the case with $u > 0$ due to symmetry.}. Denote $\boldsymbol{s}_n = [x_n,0]^\top$ as the coordinate of the \( n \)-th MA, and $r \in [r_{\text{min}}, r_{\text{max}}]$ as the distance between the origin and the target, where $r_{\text{min}}$ and $r_{\text{max}}$ are the prescribed lower-bound and upper-bound of the distance range, respectively. Thus, the coordinate of the target is given by $\boldsymbol{r} = [r \cos \theta, r \sin \theta]^\top$. Then, the distance from the \( n \)-th MA to the target can be expressed as a function of the APV $\boldsymbol{x}$ and the target parameters denoted by $\boldsymbol{\eta} = [u, r]^\top$, i.e.,
\begin{align}
\label{Taylor}
r_n(x_n, \boldsymbol{\eta}) = \lvert\boldsymbol{r}-\boldsymbol{s}_n\rvert &= \sqrt{r^2-2\boldsymbol{r}\cdot \boldsymbol{s}_n+\lvert\boldsymbol{s}_n\rvert^2} \\ 
&= \sqrt{r^2-2 r x_n u+x_n^2}. \notag
\end{align}
By invoking the Fresnel approximation for the near-field model \cite{19}, the distance in \eqref{Taylor} can be approximated as the second-order Taylor expansion of $\sqrt{1+x} \approx 1+\frac{1}{2}x-\frac{1}{8}x^2$ with $x = (-2\boldsymbol{r} \cdot \boldsymbol{s}_n+\lvert\boldsymbol{s}_n\rvert^2)/r^2$, i.e.,
\begin{align}
r_n(x_n, \boldsymbol{\eta}) &\approx r-x_n u+\frac{x_n^2(1-u^2)}{2r}.
\end{align}
Let $\tilde{\beta}$ denote the free-space path loss of the link between the origin of the MA array and the target. Then, the channel coefficient between the \( n \)-th MA and the target is given by
\begin{align}
h(x_n, \boldsymbol{\eta}) &= \sqrt{\tilde\beta} e^{-j\frac{2\pi}{\lambda}r_n(x_n, \boldsymbol{\eta})} = \beta e^{j\frac{2\pi}{\lambda}\big(x_n u-\frac{x_n^2(1-u^2)}{2r}\big)},
\end{align}
where $\beta = \sqrt{\tilde\beta} e^{-j\frac{2\pi}{\lambda}r}$ is the complex channel gain. As a result, the echoed LoS channel vector can be written as
\begin{align}
\boldsymbol{h}(\boldsymbol{x}, \boldsymbol{\eta}) &= [h(x_1, \boldsymbol{\eta}), h(x_2, \boldsymbol{\eta}), \dots, h(x_N, \boldsymbol{\eta})]^\top \\ &= \beta \boldsymbol{\alpha}(\boldsymbol{x}, \boldsymbol{\eta}) \in \mathbb{C}^N, \notag
\end{align}
where $\boldsymbol{\alpha}(\boldsymbol{x}, \boldsymbol{\eta})$ denotes the near-field steering vector of the MA array. In this paper, we aim to estimate $\boldsymbol{\eta}$ by properly setting the APV $\boldsymbol{x}$, as detailed next.

To characterize the estimation accuracy, we adopt the CRB on $\boldsymbol{\eta}$, which also serves as a theoretical lower bound on its estimation MSE. Hence, in this paper, we focus on the antenna position optimization for minimizing the CRB on $\boldsymbol{\eta}$. Note that compared to far-field sensing only involving the angular domain, near-field sensing involves both angular and spatial information, which is in favor of the target localization \cite{15}. In the following, to gain essential insights into the effects of the antenna positions on the sensing accuracy in the near-field, we consider the following three cases:
\begin{enumerate}
    \item \textbf{Estimation of AoA only (Case 1)}: $r$ is known while $u$ is unknown;
    \item \textbf{Estimation of distance only (Case 2)}: $u$ is known while $r$ is unknown;
    \item \textbf{Joint estimation of AoA and distance (Case 3)}: both $u$ and $r$ are unknown.
\end{enumerate}

\section{Antenna Position Optimization For AoA Estimation}
For the AoA estimation or Case 1, we assume that the distance from the target to the origin of the MA array is already known and denoted as $r^{\star}$, such that only the spatial AoA $u$ needs to be estimated.

\subsection{AoA Estimation}
For any given APV \(\boldsymbol{x}\), the multiple signal classification (MUSIC) algorithm can be adopted to estimate the spatial AoA \(u\) of the target. Specifically, the received echo signal at the MA array in the \(t\)-th snapshot (\(t = 1, 2, \ldots, T\)) is expressed as
\begin{equation}
    \boldsymbol{y}_t = \boldsymbol{h}(\boldsymbol{x}, u)s_t + \boldsymbol{w}_t,
\end{equation}
where \(s_t\) represents the sensing signal with \(\mathbb{E}\{\lvert s_t\rvert^2\} = P\), with $P$ denoting the transmit power, and \(\boldsymbol{w}_t \sim \mathcal{CN}(\boldsymbol{0}, \sigma^2\boldsymbol{I}_N)\) is the receiver noise following the circularly symmetric complex Gaussian (CSCG) distribution with $\sigma^2$ denoting the average noise power.

To estimate the spatial AoA \(u\), the received signals across the \(T\) snapshots are arranged into the following matrix as
\begin{equation}
\boldsymbol{Y} \triangleq
    \begin{bmatrix}
    \boldsymbol{y}_1, \boldsymbol{y}_2, \ldots, \boldsymbol{y}_T
    \end{bmatrix}
    = \boldsymbol{h}(\boldsymbol{x}, u)\boldsymbol{s}^\top + \boldsymbol{W},
\end{equation}
where \(\boldsymbol{s} \triangleq [s_1, s_2, \ldots, s_T]^\top \in \mathbb{C}^T\) and \(\boldsymbol{W} \triangleq [\boldsymbol{w}_1, \boldsymbol{w}_2, \ldots, \boldsymbol{w}_T] \in \mathbb{C}^{N \times T}\). Therefore, the covariance matrix of \(\boldsymbol{Y}\) can be given by
\begin{equation}
    \boldsymbol{R_Y} = \frac{1}{T}\boldsymbol{Y}\boldsymbol{Y}^\mathsf{H} = \frac{1}{T} \boldsymbol{h}(\boldsymbol{x}, u)\boldsymbol{s}^\mathsf{H}\boldsymbol{s}\boldsymbol{h}(\boldsymbol{x}, u)^\mathsf{H} + \sigma^2\boldsymbol{I}_N.
\end{equation}
Based on the procedures of the MUSIC algorithm, we can perform the singular value decomposition (SVD) of \(\boldsymbol{R_Y}\) as
\begin{equation}
    \boldsymbol{R_Y} = 
    \begin{bmatrix}
    \boldsymbol{u}_{\boldsymbol{s}}, \boldsymbol{U}_{\boldsymbol{w}}
    \end{bmatrix}
    \begin{bmatrix}
    \gamma_{\boldsymbol{s}} & \\
    \ & \boldsymbol{\Gamma_{\boldsymbol{w}}}
    \end{bmatrix}
    \begin{bmatrix}
    \boldsymbol{u}_{\boldsymbol{s}}^\mathsf{H} \\
    \boldsymbol{U}_{\boldsymbol{w}}^\mathsf{H}
    \end{bmatrix},
\end{equation}
where \(\boldsymbol{u}_{\boldsymbol{s}} \in \mathbb{C}^N\) and \(\boldsymbol{U}_{\boldsymbol{w}} \in \mathbb{C}^{N \times (N-1)}\) are the singular vector and matrix of the signal and noise subspaces, respectively, \(\gamma_{\boldsymbol{s}}\) denotes the singular value of the signal subspace, and \(\boldsymbol{\Gamma_{\boldsymbol{w}}} \in \mathbb{R}^{(N-1)\times(N-1)} \) represents a diagonal matrix with the singular values of the noise subspace on the diagonal. Since \( \boldsymbol{\alpha}(\boldsymbol{x}, u) \) is orthogonal to \(\boldsymbol{U}_{\boldsymbol{w}} \), and \( \boldsymbol{\alpha}(\boldsymbol{x}, \tilde{u}) \) with \( \tilde{u} \neq u \) is non-orthogonal to \( \boldsymbol{U}_{\boldsymbol{w}} \), i.e., $\boldsymbol{\alpha}(\boldsymbol{x}, u)^\mathsf{H}\boldsymbol{U}_{\boldsymbol{w}}\boldsymbol{U}_{\boldsymbol{w}}^\mathsf{H}\boldsymbol{\alpha}(\boldsymbol{x}, u) = 0 \quad \text{and} \quad \boldsymbol{\alpha}(\boldsymbol{x}, \tilde{u})^\mathsf{H}\boldsymbol{U}_{\boldsymbol{w}}\boldsymbol{U}_{\boldsymbol{w}}^\mathsf{H}\boldsymbol{\alpha}(\boldsymbol{x}, \tilde{u}) \neq 0$, there is a peak for the spectrum function $p(\Bar{u}) \triangleq \frac{1}{\boldsymbol{\alpha}(\boldsymbol{x}, \Bar{u})^\mathsf{H}\boldsymbol{U}_{\boldsymbol{w}}\boldsymbol{U}_{\boldsymbol{w}}^\mathsf{H}\boldsymbol{\alpha}(\boldsymbol{x}, \Bar{u})}$ at \( \Bar{u} = u \). Thus, the estimation of \( u \) is given by
\begin{equation}
    \hat{u} = \arg\max_{\Bar{u} \in [0, 1)} \frac{1}{\boldsymbol{\alpha}(\boldsymbol{x}, \Bar{u})^\mathsf{H} \boldsymbol{U}_{\boldsymbol{w}} \boldsymbol{U}_{\boldsymbol{w}}^\mathsf{H} \boldsymbol{\alpha}(\boldsymbol{x}, \Bar{u})},
\end{equation}
which can be solved by performing a 1D search over the peaks of the spectrum. Then, the MSE of $u$ can be expressed as $\text{MSE}(u) \triangleq \mathbb{E}\{\lvert u - \hat{u}\rvert^2\}$, and its CRB is given by \cite{22, 23}
\begin{equation}
    \text{CRB}_u(\boldsymbol{x}, u) = \frac{\kappa}{F_u(\boldsymbol{x}, u)} \le \text{MSE}(u),
\label{CRBu}
\end{equation}
where $\kappa \triangleq \frac{\sigma^2 \lambda^2}{8\pi^2 T P N \lvert\beta\rvert^2}$ and
\begin{equation}
    F_u(\boldsymbol{x}, u) \triangleq \text{var}(\boldsymbol{x})+\frac{2u}{r^\star}\text{cov}(\boldsymbol{x},\boldsymbol{\tilde{x}})+\frac{u^2}{{r^\star}^2}\text{var}(\boldsymbol{\tilde{x}}),
\label{Fxu}
\end{equation}
where $\tilde{\boldsymbol{x}} \triangleq [\tilde{x}_1, \tilde{x}_2, \dots, \tilde{x}_N]^\top \in \mathbb{R}^N$ and $\tilde{x}_n \triangleq x_n^2, n \in {\cal N}$. The variance functions are defined as $\text{var}(\boldsymbol{x}) \triangleq \frac{1}{N} \sum_{n=1}^N x_n^2 - \mu(\boldsymbol{x})^2$ with $\mu(\boldsymbol{x}) = \frac{1}{N} \sum_{n=1}^N x_n$ being the mean of \(\boldsymbol{x}\), and $\text{var}(\tilde{\boldsymbol{x}}) \triangleq \frac{1}{N} \sum_{n=1}^N \tilde{x}_n^2 - \mu(\tilde{\boldsymbol{x}})^2$ with $\mu(\tilde{\boldsymbol{x}}) = \frac{1}{N} \sum_{n=1}^N \tilde{x}_n$ being the mean of \(\tilde{\boldsymbol{x}}\). The covariance function is defined as $\text{cov}(\boldsymbol{x},\tilde{\boldsymbol{x}}) \triangleq \frac{1}{N} \sum_{n=1}^N x_n \tilde{x}_n - \mu(\boldsymbol{x})\mu(\tilde{\boldsymbol{x}})$.

\subsection{Problem Formulation and Proposed Solution}
Our objective is to minimize $\text{CRB}_u(\boldsymbol{x}, u)$ by optimizing the APV $\boldsymbol{x}$. However, the CRB in \eqref{CRBu} is dependent on both the APV \(\boldsymbol{x}\) and AoA itself. To tackle this issue, we focus on minimizing the worst-case $\text{CRB}_u(\boldsymbol{x}, u)$ for all possible values of $u$, i.e., $\max_{u} \text{CRB}_u(\boldsymbol{x}, u)$. The associated min-max problem can be easily shown equivalent to the following max-min problem based on \eqref{CRBu}, i.e.,
\begin{equation}
    \min_{\boldsymbol{x}} \max_{u\in[0, 1)} \text{CRB}_u(\boldsymbol{x}, u) \iff \max_{\boldsymbol{x}} \min_{u\in[0, 1)} F_u(\boldsymbol{x}, u).
\label{minmax}
\end{equation}
The associated optimization problem for the right-hand side of \eqref{minmax} can be formulated as
\begin{subequations}
\label{P1}
\begin{align}
\text{(P1)} \quad
& \max_{\boldsymbol{x}} \; F_u^\star(\boldsymbol{x}) \triangleq \text{var}(\boldsymbol{x})+\frac{2u_{\text{opt}}}{r^\star}\text{cov}(\boldsymbol{x},\boldsymbol{\tilde{x}})+\frac{u^2_{\text{opt}}}{{r^\star}^2}\text{var}(\boldsymbol{\tilde{x}}) \label{P1a} \\
& \text{s.t.} \quad 0 \le x_n \leq A, \quad n \in {\cal N}, \label{P1b} \\
& \phantom{\text{s.t.} \quad} \lvert x_n - x_{n-1} \rvert \geq d, \quad n \in {\cal N}\backslash \{1\}, \label{P1c}
\end{align}
\end{subequations}
where $d$ denotes the minimum inter-MA distance to avoid mutual coupling, and $u_{\text{opt}}$ is the AoA value that yields the worst-case CRB on the AoA, i.e., $u_{\text{opt}}= \arg\max_{u} \text{CRB}_u(\boldsymbol{x}, u)$. Note that since $u_{\text{opt}} \geq 0$, it is desirable that the MAs should be positioned as dispersive apart as possible to maximize $F_u^\star(\boldsymbol{x})$, which helps increase the variance terms $\text{var}(\boldsymbol{x})$ and $\text{var}(\tilde{\boldsymbol{x}})$, as well as the covariance term $\text{cov}(\boldsymbol{x},\tilde{\boldsymbol{x}})$. This will be rigorously proved in Theorem \ref{Th1} below.
\begin{figure}[t]
    \centering
    \includegraphics[scale=0.4]{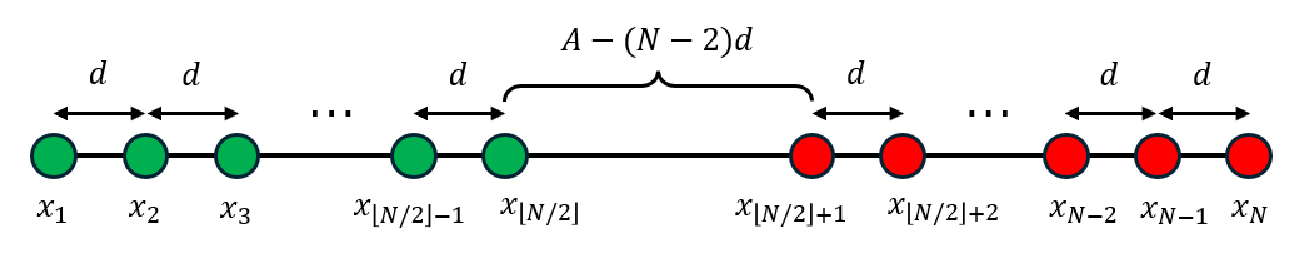}
    \caption{\small The optimal APV in the AoA or distance estimation.}
    \label{Fig2}
\end{figure}
\begin{theorem} \label{Th1}
The optimal solution to (P1) is given by
\begin{equation}
x_n^\star = 
\begin{cases} 
(n-1)d, & n = 1, 2, \ldots, \lfloor N/2 \rfloor; \\
A - (N-n)d, & n = \lfloor N/2 \rfloor + 1, \ldots, N.
\end{cases}
\end{equation}
\end{theorem}
Theorem \ref{Th1} can be proved via a similar process to Appendix A in \cite{13}, for which the detailed proof is omitted due to the page limit. It demonstrates that, to minimize the CRB of the AoA estimation MSE, the optimal MA positions are the same as those in the AoA estimation in the far-field, as derived in \cite{13}. In particular, the MAs should be divided into two groups, as depicted in Fig. \ref{Fig2}. The first group of MAs is placed at the leftmost end of the 1D line segment, while the other group at the rightmost end. Additionally, it can be shown that the CRB on the AoA can be effectively decreased by increasing the length of the line segment, as this results in a larger array aperture, enabling the synthesis of sensing beams with higher angular resolution in the near-field region for a given distance.

\section{Antenna Position Optimization for Distance Estimation}
In this section, we consider Case 2 where the spatial AoA is already known, denoted as $u^{\star}$. To estimate the distance $r$, we also apply the MUSIC algorithm by leveraging the distance-related information in the signal phase \cite{18}. For simplicity, the detailed process for distance estimation is omitted here. The associated MSE and CRB are given by
\begin{equation}
    \text{MSE}(r) \geq \text{CRB}_r(\boldsymbol{x}, r) = \kappa \cdot F_r^{-1}(\boldsymbol{x}, r),
\label{CRBr}
\end{equation}
where
\begin{equation}
    F_r(\boldsymbol{x}, r) \triangleq \Big(\frac{1-{u^\star}^2}{2r^2}\Big)^2 \text{var}(\tilde{\boldsymbol{x}}).
\label{Fxr}
\end{equation}
Note that the CRB in \eqref{CRBr} depends on the distance $r$. To eliminate its effects, similar to the AoA estimation, we aim to minimize the maximum of the right-hand side of \eqref{CRBr} among all possible values of $r$, i.e., $\max_{r} \text{CRB}_r(\boldsymbol{x}, r)$. Hence, the corresponding min-max problem can be reformulated as a max-min problem, i.e.,
\begin{equation}
    \min_{\boldsymbol{x}} \max_{r \in [r_{\text{min}}, r_{\text{max}}]} \text{CRB}_r(\boldsymbol{x}, r) \iff \max_{\boldsymbol{x}} \min_{r \in [r_{\text{min}}, r_{\text{max}}]} F_r(\boldsymbol{x}, r).
\end{equation}
Then, the optimization problem can be formulated as
\begin{subequations}
\label{P2}
\begin{align}
\text{(P2)} \quad
& \max_{\boldsymbol{x}} \quad F_r^\star(\boldsymbol{x}) \triangleq \Big(\frac{1-{u^\star}^2}{2r^2_{\text{opt}}}\Big)^2 \text{var}(\tilde{\boldsymbol{x}}) \\
& \text{s.t.} \quad \eqref{P1b}, \eqref{P1c}, \notag
\end{align}
\end{subequations}
where $r_{\text{opt}}$ is the distance value that yields the worst-case CRB on the distance, i.e., $r_{\text{opt}}= \arg\max_{r} \text{CRB}_r(\boldsymbol{x}, r)$. It can be easily proved that the optimal APV to (P2) the same as that provided in Theorem \ref{Th1} via a similar process to Appendix A in \cite{13}. As a result, in the individual estimation of $u$ or $r$, the associated optimal APVs are identical and the same as the AoA estimation in the far-field sensing. The optimal MA positions should maximally increase the aperture to ensure the sensing resolution in the angle or distance domain.



\section{Antenna Position Optimization for Joint AoA and Distance Estimation}
In the joint estimation of the spatial AoA and distance or Case 3, we implement the 2D-MUSIC algorithm that performs an exhaustive search on the two-dimensional (2D) angle-distance grid to find the peaks of the 2D spectrum function. This process allows us to estimate both the angle and the distance \cite{18}. Therefore, the joint estimation result is given by
\begin{equation}
    \hat{\boldsymbol{\eta}} = \arg\max_{\Bar{\boldsymbol{\eta}} \in [0, 1) \times [r_{\text{min}}, r_{\text{max}}]} \frac{1}{\boldsymbol{\alpha}(\boldsymbol{x}, \Bar{\boldsymbol{\eta}})^\mathsf{H} \boldsymbol{U}_{\boldsymbol{w}} \boldsymbol{U}_{\boldsymbol{w}}^\mathsf{H} \boldsymbol{\alpha}(\boldsymbol{x}, \Bar{\boldsymbol{\eta}})}.
\end{equation}
Accordingly, we aim to minimize the CRBs on the joint estimation MSEs of AoA and distance by optimizing the MA positions. To this end, we first calculate the Fisher information matrix (FIM) of the estimator. Based on the 2D-MUSIC algorithm, the FIM of the estimator $\boldsymbol{\eta}$ can be expressed as
\begin{equation}
    \text{FIM}(\boldsymbol{x}, \boldsymbol{\eta}) = 
    \begin{bmatrix}
    J_{uu} & J_{ur} \\
    J_{ru} & J_{rr} 
    \end{bmatrix},
\label{FIMeta}
\end{equation}
where the detailed derivations of the FIMs $J_{\alpha \beta} (\alpha, \beta \in \{u, r\})$ are given in Appendix. Then, the CRB matrix is derived by taking the inverse of the FIM of the estimator, i.e., $\text{CRB}_{\boldsymbol{\eta}} (\boldsymbol{x}, \boldsymbol{\eta}) = \text{FIM}^{-1}(\boldsymbol{x}, \boldsymbol{\eta})$. Specifically, the CRBs on AoA and distance in the joint estimation are given by
\begin{equation}
    \text{CRB}_u(\boldsymbol{x}) = \kappa \cdot \text{var}(\boldsymbol{\tilde{x}})\big/\big(\text{var}(\boldsymbol{x})\text{var}(\boldsymbol{\tilde{x}})-{\text{cov}^2(\boldsymbol{x},\boldsymbol{\tilde{x}})}\big),
\label{CRBjoint_u}
\end{equation}
\begin{equation}
    \text{CRB}_r(\boldsymbol{x}, \boldsymbol{\eta}) = \frac{\kappa\Big(4r^4\text{var}(\boldsymbol{x})+8ur^3{\text{cov}(\boldsymbol{x},\boldsymbol{\tilde{x}})}+4u^2r^2\text{var}(\boldsymbol{\tilde{x}})\Big)}{(1-u^2)^2\Big(\text{var}(\boldsymbol{x})\text{var}(\boldsymbol{\tilde{x}})-\text{cov}^2(\boldsymbol{x},\boldsymbol{\tilde{x}})\Big)},
\label{CRBjoint_r}
\end{equation}
respectively. The detailed procedure for deriving the CRBs is given in Appendix as well. It can be observed from \eqref{CRBjoint_u} and \eqref{CRBjoint_r} that there may exist a fundamental trade-off between minimizing $\text{CRB}_u(\boldsymbol{x})$ and $\text{CRB}_r(\boldsymbol{x}, \boldsymbol{\eta})$ due to the complicated coupling between the variance and covariance terms therein. Moreover, although $\text{CRB}_u(\boldsymbol{x})$ is independent of the spatial AoA $u$, $\text{CRB}_r(\boldsymbol{x}, \boldsymbol{\eta})$ depends on both the spatial AoA $u$ and distance $r$. To overcome this difficulty, we aim to minimize the sum of $\text{CRB}_u(\boldsymbol{x})$ and the worst-case $\text{CRB}_r(\boldsymbol{x}, \boldsymbol{\eta})$, i.e., $\text{CRB}_u(\boldsymbol{x})+\max_{u, r}\text{CRB}_r(\boldsymbol{x}, \boldsymbol{\eta})$, by optimizing the APV $\boldsymbol{x}$. Denote $\boldsymbol{\eta}_{\text{opt}}$ as the estimator vector that yields the worst-case CRB on the distance. Then, the associated optimization problem can be equivalently transformed into the following problem, i.e.,
\begin{subequations}
\label{P3}
\begin{align}
\text{(P3)} \quad 
& \max_{\boldsymbol{x}} \quad F_{\boldsymbol{\eta}}^\star(\boldsymbol{x}) \triangleq\text{Tr}^{-1}\Big(\text{CRB}_{\boldsymbol{\eta}} (\boldsymbol{x}, \boldsymbol{\eta})\Big)\Big|_{\boldsymbol{\eta}=\boldsymbol{\eta}_{\text{opt}}} \label{P3a} \\
& \text{s.t.} \quad \eqref{P1b}, \eqref{P1c}. \notag
\end{align}
\end{subequations}
However, it can be observed from \eqref{CRBjoint_u} and \eqref{CRBjoint_r} that the objective function of (P3) is non-convex w.r.t. the APV $\boldsymbol{x}$, which is challenging to be optimally solved. Therefore, we utilize a discrete sampling-based algorithm \cite{9, 12} to derive a high-quality sub-optimal APV solution denoted as $\boldsymbol{x}^\star$, by sequentially selecting the optimal sampling points for MAs.
\begin{algorithm}[t]
\caption{Proposed Algorithm for Solving Problem (P3)}
\label{discrete}
\begin{algorithmic}[1]
\State \textbf{Input:} $n=1$, ${\cal X}$, and ${\cal S}_1$.
\While{$n \leq N$}
    \State Obtain $x^\star_n$ based on \eqref{arg} and update $x^{\text{init}}_n \gets x^\star_n$.
    \State Determine ${\cal S}_{n+1}$ based on \eqref{set}.
    \State Update $n \gets n+1$.
\EndWhile
\State \textbf{Output:} the optimized APV of all $N$ MAs, i.e., $\boldsymbol{x}^\star$.
\end{algorithmic}
\end{algorithm}

Specifically, the continuous MA array is uniformly discretized into $M$ $(M \gg N)$ sampling points, with the distance between any two adjacent sampling points denoted by $\delta_s =A/M$ and the position of the $i$-th sampling point given by $x_i=i \delta_s, i \in {\cal M} \triangleq \{1, 2, \ldots, M\}$. By denoting ${\cal S}=\{x_i | i \in {\cal M}\}$ as the set of all sampling points, we first construct an initial set of the positions of the $N$ MAs, denoted by ${\cal X}=\{x^{\text{init}}_n | x^{\text{init}}_n \in {\cal S}, n \in {\cal N}\}$. In the $n$-th iteration, we only update the position of the $n$-th MA, i.e., $x^{\text{init}}_n$, while keeping the positions of other $(N-1)$ MAs fixed. Let $x^\star_n$ denote the updated position of the $n$-th MA in the $n$-th iteration. Hence, the set of all feasible sampling points for updating $x^{\text{init}}_n$ is
\begin{align}
     {\cal S}_n = \{s | s \in {\cal S}, & \lvert s-x_i^\star \rvert \geq d, 1 \leq i \leq n-1, \lvert s-x^{\text{init}}_j \rvert \geq d, \notag \\
     & n+1 \leq j \leq N\}, 2 \leq n \leq N-1,
\label{set}
\end{align}
and we set ${\cal S}_1 = \{s | s \in {\cal S}, \lvert s-x^{\text{init}}_j \rvert \geq d, 2 \leq j \leq N\}$ and ${\cal S}_N = \{s | s \in {\cal S}, \lvert s-x_i^\star \rvert \geq d, 1 \leq i \leq N-1\}$. Then, we update $x^{\text{init}}_n$ as $x_n^\star$ by maximizing the objective function in (P3), i.e.,
\begin{equation}
    x_n^\star = \arg\max_{s \in {\cal S}_n} F_{\boldsymbol{\eta}}^\star(\boldsymbol{\hat{x}}_{n}),
\label{arg}
\end{equation}
where $\boldsymbol{\hat{x}}_{n}=[x^\star_1,\ldots,x^\star_{n-1},s,x^{\text{init}}_{n+1},\ldots,x^{\text{init}}_N]^\top$. Next, in the $(n+1)$-th iteration, we proceed to update ${\cal S}_{n+1}$ based on \eqref{set} and then update the $(n+1)$-th MA position based on \eqref{arg}. Note that the above sequential update process can yield a non-decreasing objective function value of (P3); hence, its convergence is guaranteed. The main procedures of the proposed algorithm for solving problem (P3) are summarized in Algorithm \ref{discrete}. Furthermore, it can be concluded that the computational complexity of Algorithm \ref{discrete} is ${\cal O}(NM)$.
\begin{figure}[t]
    \centering
    \includegraphics[scale=0.38]{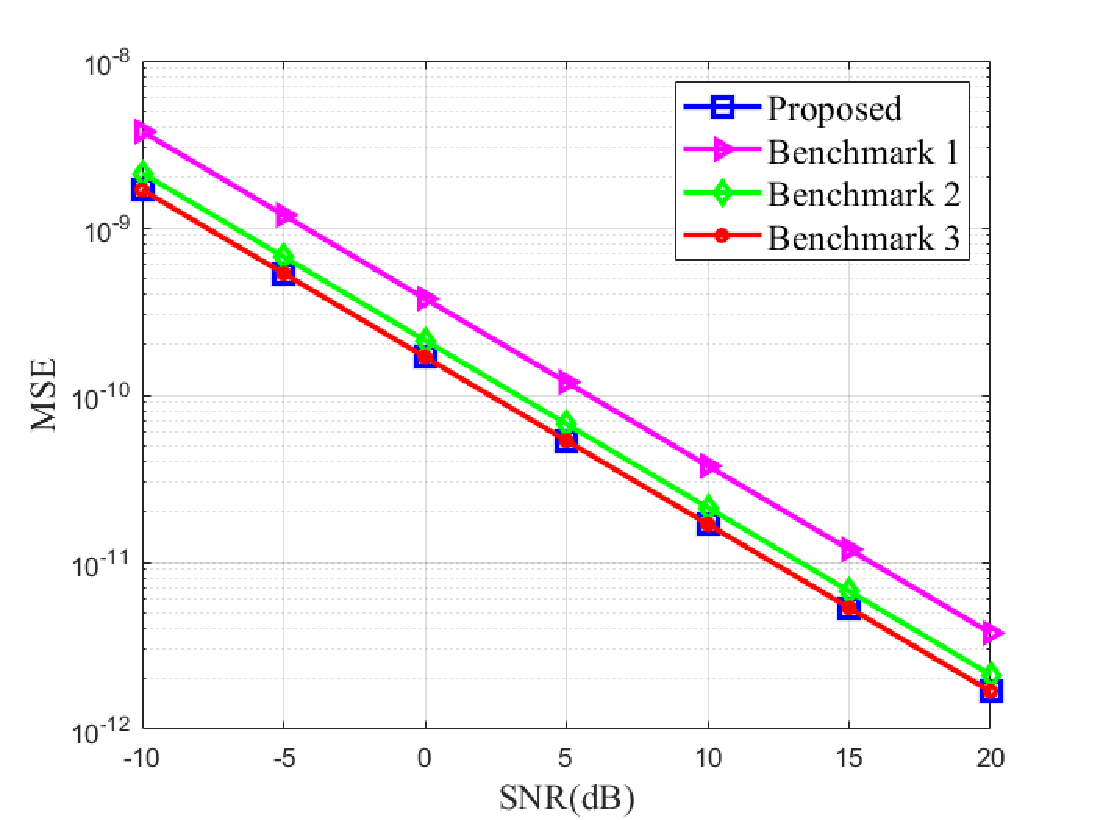}
    \caption{\small $\text{CRB}$ of the AoA estimation MSE versus the received SNR.}
    \label{case1}
\end{figure}
\begin{figure}[t]            
    \centering
    \includegraphics[scale=0.38]{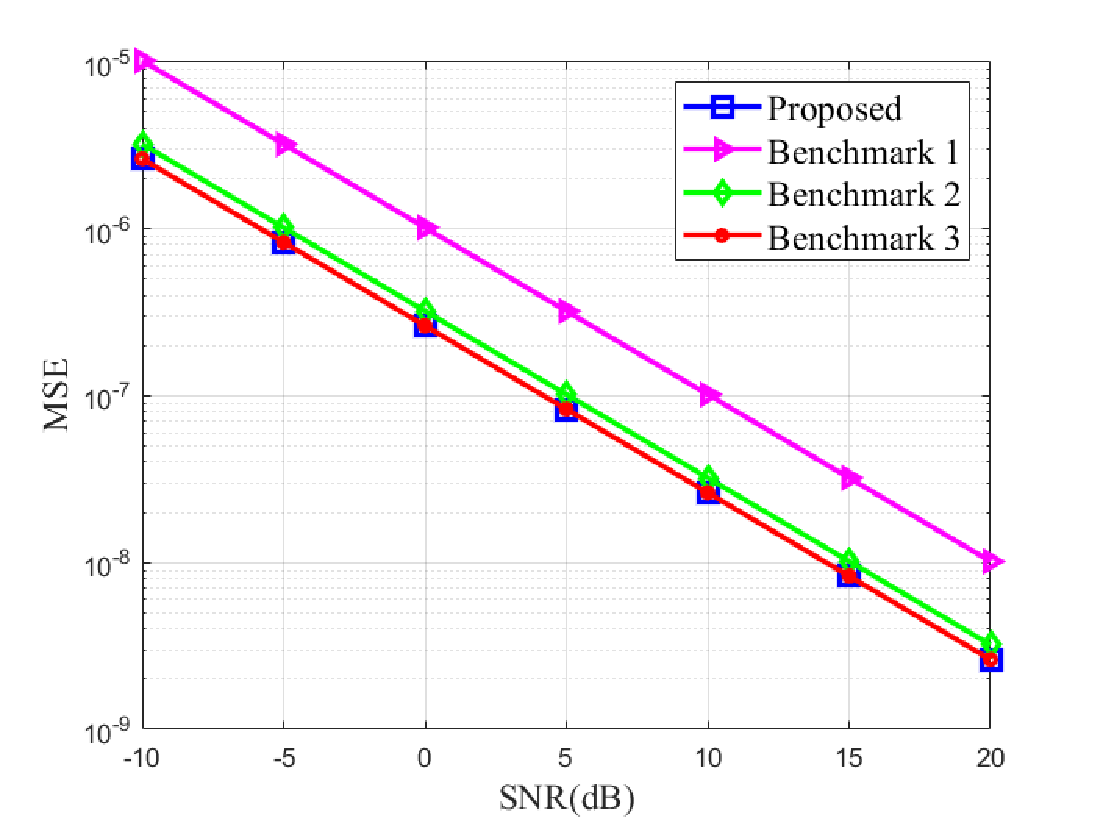}
    \caption{\small $\text{CRB}$ of the distance estimation MSE versus the received SNR.}
    \label{case2}
\end{figure}
\begin{figure}[t]
    \centering
    \includegraphics[scale=0.4]{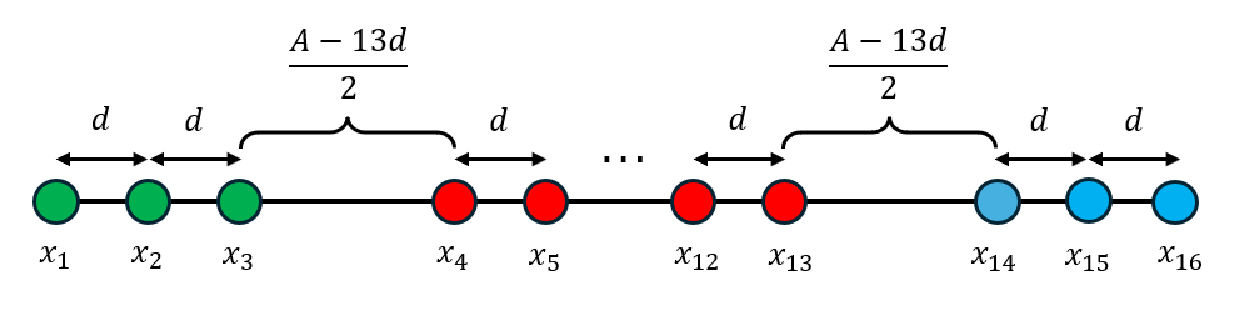}
    \caption{\small The optimal APV in the joint estimation of AoA and distance.}
    \label{case3_APV}
\end{figure}
\begin{figure}[t]
    \centering
    \includegraphics[scale=0.38]{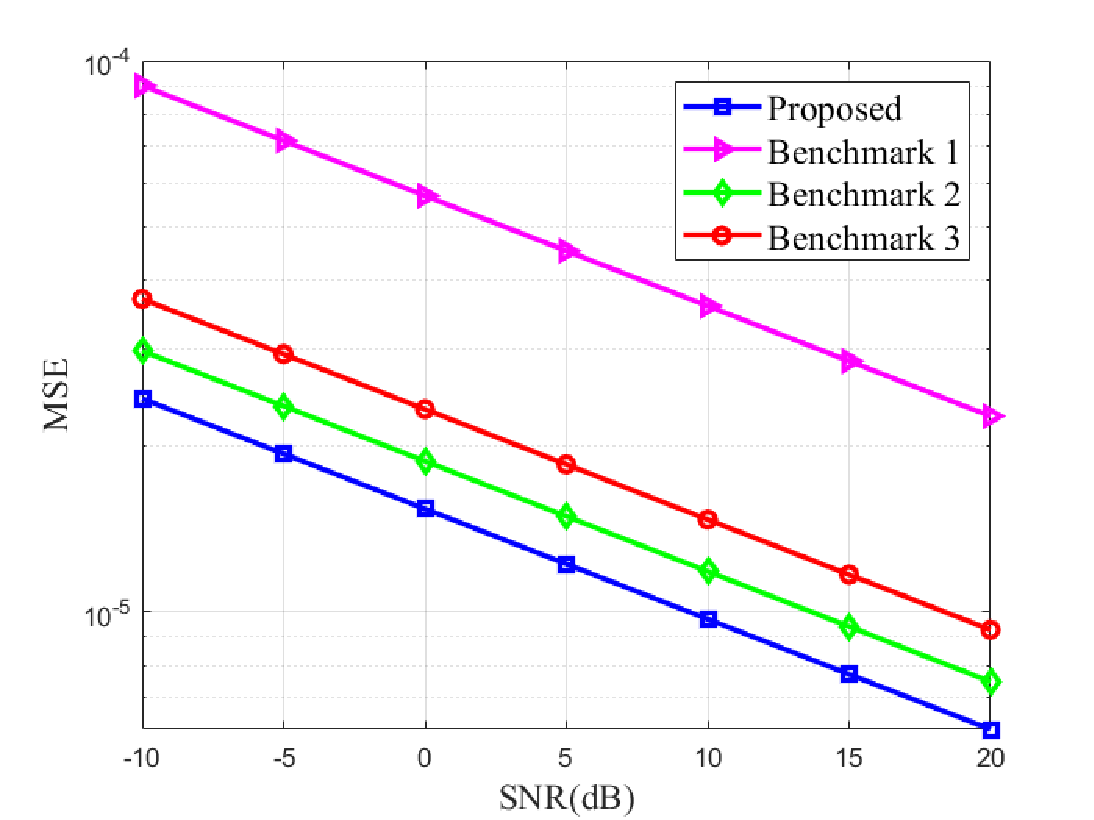}
    \caption{\small Sum of the joint estimation CRBs versus the received SNR.}
    \label{case3}
\end{figure}

\section{Numerical Results}
\label{numerical}
In this section, numerical results are presented to assess the performance of the proposed near-field sensing scheme with MAs. We set the number of Tx antennas to $N = 16$. The minimum separation between adjacent antennas and the length of the MA array are set to $d = \lambda/2$ and $A = 10\lambda$, respectively. The average received SNR is defined as $P \lvert\beta\rvert^2 / \sigma^2$. In addition, the number of snapshots is set as $T = 1$. In the AoA estimation, we set $r^\star=R_{RL}/4$. In the distance estimation, we set $\theta=45^\circ$, i.e., $u^\star=\cos\theta=0.71$. Futhermore, we set $r_{\text{min}} = R_{FS}$ and $r_{\text{max}} = R_{RL}/2$, i.e., $r \in [R_{FS}, R_{RL}/2]$ in the distance and joint estimation. In Algorithm \ref{discrete}, we set the number of sampling points to $M=2000$ and the initial APV $\boldsymbol{x}^{\text{init}}$ the same as the APV provided in Theorem \ref{Th1}. For comparison, we consider the following three benchmarks:
\begin{enumerate}
    \item \textbf{Uniform linear array (ULA) with half-wavelength antenna spacing (Benchmark 1)}: $\{x_n\}_{n=1}^N$ are set as $x_n=(n-1)d,n \in {\cal N}$;
    \item \textbf{Sparse ULA with a full aperture (Benchmark 2)}: $\{x_n\}_{n=1}^N$ are set as $x_n=(n-1)A/(N-1),n \in {\cal N}$;
    \item \textbf{Optimal MA array in the far-field AoA estimation (Benchmark 3) \cite{13}}: $\{x_n\}_{n=1}^N$ are set the same as those provided in Theorem \ref{Th1}.
\end{enumerate}

In Fig. \ref{case1}, we show the CRB of the AoA estimation MSE in \eqref{CRBu} versus the received SNR by different schemes. It is observed that the proposed optimal APV in Theorem \ref{Th1} results in a significantly lower CRB compared to Benchmarks 1 and 2. For SNR = 20 dB, the optimal APV is observed to yield 55.3\% and 20.5\% CRB reduction over Benchmarks 1 and 2, respectively. Benchmark 1 is observed to achieve the worst performance among all considered schemes, as its effective aperture is the smallest, resulting in limited angle resolution. It is also observed that although Benchmark 2 has the same effective aperture as the proposed scheme, the array geometry may not minimize the CRB, thus resulting in a worse performance than ours. Moreover, Benchmark 3 is observed to achieve the same performance as the proposed scheme, as expected. For the distance estimation, the CRBs of its MSEs in \eqref{CRBr} versus the received SNR by different schemes are shown in Fig. \ref{case2}. Similar observations made from Fig. \ref{case1} can also be made in Fig. \ref{case2}. Particularly, compared to Benchmarks 1 and 2, the proposed scheme leads to a notable decrease in the CRB. For SNR = 20 dB, the CRB is reduced by 74.2\% and 18.4\% over Benchmarks 1 and 2, respectively.

In Fig. \ref{case3_APV}, we show the optimized positions of the MAs for the joint estimation of AoA and distance. It is observed that unlike the array geometry shown in Fig. \ref{Fig2} for AoA/distance estimation only, that for the joint estimation consists of three groups of antennas, as marked by different colors. In each group, any two adjacent MAs are spaced by half-wavelength, and the spacing between the leftmost/rightmost group and the middle group is identical. Moreover, the first and the $N$-th MAs are placed at the two endpoints of the array, respectively, i.e., $x_1=0$ and $x_N=A$, which ensures the maximum array aperture to increase the estimation resolution. To verify the effectiveness of the proposed scheme, we show the worst-case sum of the CRBs in \eqref{CRBjoint_u} and \eqref{CRBjoint_r} versus the received SNR in Fig. \ref{case3}. It is seen that the optimal APV results in a remarkable decrease in the CRB compared to the three benchmark schemes. Specifically, for SNR = 20 dB, the proposed scheme achieves 73.0\%, 34.0\% and 18.1\% reduction over Benchmarks 1, 3, and 2, respectively. It is interesting to note that different from the observations made from Figs. \ref{case1} and \ref{case2}, Benchmark 3 is observed to yield a much worse performance than the proposed scheme. The possible reason is that the three-group geometry shown in Fig. \ref{case3_APV} can enable the antenna array to more accurately sense the target in proximity to its center.

\section{Conclusion}
In this paper, we investigated MA-enhanced near-field sensing, aiming to estimate the spatial AoA or/and distance information of a target. By adopting the MUSIC algorithm, we derived the worst-case CRBs for the spatial AoA or/and distance estimation in different cases and optimized the MA positions to minimize them. Numerical results demonstrated that using MAs can substantially decrease the CRBs compared to conventional FPAs. In addition, even with the same antenna aperture, the optimal MA array geometry may vary for different scenarios, e.g., near- versus far-field, individual versus joint estimation.

\section*{Appendix}
\section*{Derivations of the FIM and CRB matrix}
In the joint estimation of the spatial AoA and distance of a target, the FIM of the estimator $\boldsymbol{\eta}$ in \eqref{FIMeta} is given by \cite{22, 23}
\begin{multline}
    \text{FIM}(\boldsymbol{x}, \boldsymbol{\eta}) = \frac{2}{\sigma^2}\sum_{t=1}^{T}\Re\Bigg\{s_t^* \boldsymbol{\Psi}(\boldsymbol{x}, \boldsymbol{\eta})^\mathsf{H} \bigg(\boldsymbol{I}_N-\boldsymbol{\alpha}(\boldsymbol{x}, \boldsymbol{\eta}) \\
    \Big(\boldsymbol{\alpha}(\boldsymbol{x}, \boldsymbol{\eta})^\mathsf{H} \boldsymbol{\alpha}(\boldsymbol{x}, \boldsymbol{\eta})\Big)^{-1}\boldsymbol{\alpha}(\boldsymbol{x}, \boldsymbol{\eta})^\mathsf{H}\bigg)\boldsymbol{\Psi}(\boldsymbol{x}, \boldsymbol{\eta})s_t\Bigg\},
\label{app_FIM}
\end{multline}
where $\boldsymbol{\Psi}(\boldsymbol{x}, \boldsymbol{\eta})$ denotes the partial derivative matrix of the near-field steering vector $\boldsymbol{\alpha}(\boldsymbol{x}, \boldsymbol{\eta})$ w.r.t. the estimator $\boldsymbol{\eta}$, i.e.,
\begin{equation}
    \boldsymbol{\Psi}(\boldsymbol{x}, \boldsymbol{\eta}) = \Big[\frac{\partial \boldsymbol{\alpha}(\boldsymbol{x}, \boldsymbol{\eta})}{\partial u}, \frac{\partial \boldsymbol{\alpha}(\boldsymbol{x}, \boldsymbol{\eta})}{\partial r}\Big] \in \mathbb{C}^{N \times 2}.
\end{equation}
Hence, by \eqref{app_FIM}, the four elements of $\text{FIM}(\boldsymbol{x}, \boldsymbol{\eta})$ are given by
\begin{equation}
    J_{uu} = \kappa^{-1} \cdot \Big(\text{var}(\boldsymbol{x})+\frac{2u}{r}\text{cov}(\boldsymbol{x},\boldsymbol{\tilde{x}})+\frac{u^2}{r^2}\text{var}(\boldsymbol{\tilde{x}})\Big),
\label{Juu}
\end{equation}
\begin{equation}
    J_{ur} = J_{ru} = \kappa^{-1} \cdot \frac{1-u^2}{2r^2}\Big(\text{cov}(\boldsymbol{x},\boldsymbol{\tilde{x}})+\frac{u}{r}\text{var}(\boldsymbol{\tilde{x}})\Big),
\label{Jur}
\end{equation}
\begin{equation}
    J_{rr} = \kappa^{-1} \cdot \Big(\frac{1-u^2}{2r^2}\Big)^2\text{var}(\boldsymbol{\tilde{x}}),
\label{Jrr}
\end{equation}
respectively. Then, the CRB matrix for the joint estimation is obtained by taking the inverse of $\text{FIM}(\boldsymbol{x}, \boldsymbol{\eta})$, i.e.,
\begin{equation}
\text{CRB}_{\boldsymbol{\eta}} (\boldsymbol{x}, \boldsymbol{\eta})=
    \begin{bmatrix}
    \Big(J_{uu}-\dfrac{J_{ur}J_{ru}}{J_{rr}}\Big)^{-1} & \Big(J_{ru}-\dfrac{J_{uu}J_{rr}}{J_{ur}}\Big)^{-1} \\[8pt]
    \Big(J_{ur}-\dfrac{J_{uu}J_{rr}}{J_{ru}}\Big)^{-1} & \Big(J_{rr}-\dfrac{J_{ru}J_{ur}}{J_{uu}}\Big)^{-1}
    \end{bmatrix}.
\end{equation}
Therefore, the CRBs on the AoA in \eqref{CRBjoint_u} and the distance in \eqref{CRBjoint_r} are respectively given by
\begin{equation}
    \text{CRB}_u(\boldsymbol{x})=(J_{uu}-J_{ur}J_{ru}/J_{rr})^{-1},
\label{app_CRBu}
\end{equation}
\begin{equation}
    \text{CRB}_r(\boldsymbol{x}, \boldsymbol{\eta})=(J_{rr}-J_{ru}J_{ur}/J_{uu})^{-1}.
\label{app_CRBr}
\end{equation}
By substituting \eqref{Juu}, \eqref{Jur} and \eqref{Jrr} into \eqref{app_CRBu} and \eqref{app_CRBr}, we can obtain \eqref{CRBjoint_u} and \eqref{CRBjoint_r}.



\end{document}